\documentclass[11pt,
twocolumn,
 aip,
 apl,%
 showkeys,
 amsmath,amssymb,
reprint
]{revtex4-1}
\usepackage[breaklinks=true,colorlinks=true,linkcolor=blue,urlcolor=blue,citecolor=blue]{hyperref}
\usepackage{sidecap}
\usepackage{graphicx}%
\usepackage{graphics}%
\usepackage{epstopdf}
\usepackage{dcolumn}%
\usepackage{soul,color}
\usepackage{bm}
\usepackage[mathlines]{lineno}
\expandafter\ifx\csname package@font\endcsname\relax\else
 \expandafter\expandafter
 \expandafter\usepackage
 \expandafter\expandafter
 \expandafter{\csname package@font\endcsname}%
\fi
\usepackage[utf8]{inputenc}
\usepackage[T1]{fontenc}
\usepackage{mathptmx} 

\begin{document}
\abovedisplayskip=3pt
\belowdisplayskip=3pt
\abovedisplayshortskip=2pt
\belowdisplayshortskip=2pt


\title{\Large{Steering of vortices by magnetic-field tilting in superconductor nanotubes
}}
\author{Igor Bogush}
    \affiliation{Institute for Emerging Electronic Technologies, Leibniz IFW Dresden, Helmholtzstra{\ss}e 20, 01069 Dresden, Germany}
    \affiliation{Moldova State University, Str. A. Mateevici 60, 2009 Chi\c{s}in\u{a}u, Republic of Moldova}
\author{Oleksandr V. Dobrovolskiy}
    \affiliation{University of Vienna, Faculty of Physics, Nanomagnetism and Magnonics, Superconductivity and Spintronics Laboratory, W\"ahringer Str. 17, 1090 Vienna, Austria}
\author{Vladimir M. Fomin}
    \email{Vladimir M. Fomin, v.fomin@ifw-dresden.de}
    \affiliation{Institute for Emerging Electronic Technologies, Leibniz IFW Dresden, Helmholtzstra{\ss}e 20, 01069 Dresden, Germany}
    \affiliation{Moldova State University, Str. A. Mateevici 60, 2009 Chi\c{s}in\u{a}u, Republic of Moldova}
\date{\today}

\begin{abstract}
In planar superconductor thin films, the places of nucleation and arrangements of moving vortices are determined by structural defects. However, various applications of superconductors require reconfigurable steering of fluxons, which is hard to realize with geometrically predefined vortex pinning landscapes. Here, on the basis of the time-dependent Ginzburg-Landau equation, we present an approach for steering of vortex chains and vortex jets in superconductor nanotubes containing a slit. The idea is based on tilting of the magnetic field $\mathbf{B}$ at an angle $\alpha$ in the plane perpendicular to the axis of a nanotube carrying an azimuthal transport current. Namely, while at $\alpha=0^\circ$ vortices move paraxially in opposite directions within each half-tube, an increase of $\alpha$ displaces the areas with the close-to-maximum normal component $|B_\mathrm{n}|$ to the close(opposite)-to-slit regions, giving rise to descending (ascending) branches in the induced-voltage frequency spectrum $f_\mathrm{U}(\alpha)$. At lower $B$, upon reaching the critical angle $\alpha_\mathrm{c}$, close-to-slit vortex chains disappear, yielding $f_\mathrm{U}$ of the $nf_1$-type ($n\geq1$: an integer; $f_1$: vortex nucleation frequency). At higher $B$, $f_\mathrm{U}$ is largely blurry because of multifurcations of vortex trajectories, leading to the coexistence of a vortex jet with two vortex chains at $\alpha=90^\circ$. In addition to prospects for tuning of GHz-frequency spectra and steering of vortices as information bits, our findings lay foundations for on-demand tuning of vortex arrangements in 3D superconductor membranes in tilted magnetic fields.
\end{abstract}
\maketitle

Knowledge of how magnetic flux quanta (Abrikosov vortices) move and arrange themselves under various currents and magnetic fields is critical for supercurrent flow and fluxonic applications. For instance, while for defect-free planar thin films a hexagonal vortex lattice\,\cite{Bra95rpp} is expected, sample defects and its geometry make vortex patterns to differ from a regular lattice\,\cite{Rei97prl,Gui09nph,Sil10prl}. In planar thin films, the places of nucleation of vortices are determined by edge defects\,\cite{Cer13njp,Bud22pra,Vod15sst}, current-crowding effects\,\cite{Cle11prb,Ada13apl,Emb17nac} or their combinations\,\cite{Fri01prb,Ust22etp,Bev23pra}. For structures with perfect edges, a single edge defect acts as a local injector of vortices\,\cite{Ala01pcs}. Driven by the competing current-vortex and vortex-vortex interactions, such vortices form a jet which is narrow at the defect and \emph{expands} due to the repulsion of vortices as they move to the opposite edge of the structure\,\cite{Bez22prb}.

\begin{figure}[b!]
    \centering
    \includegraphics[width=0.95\linewidth]{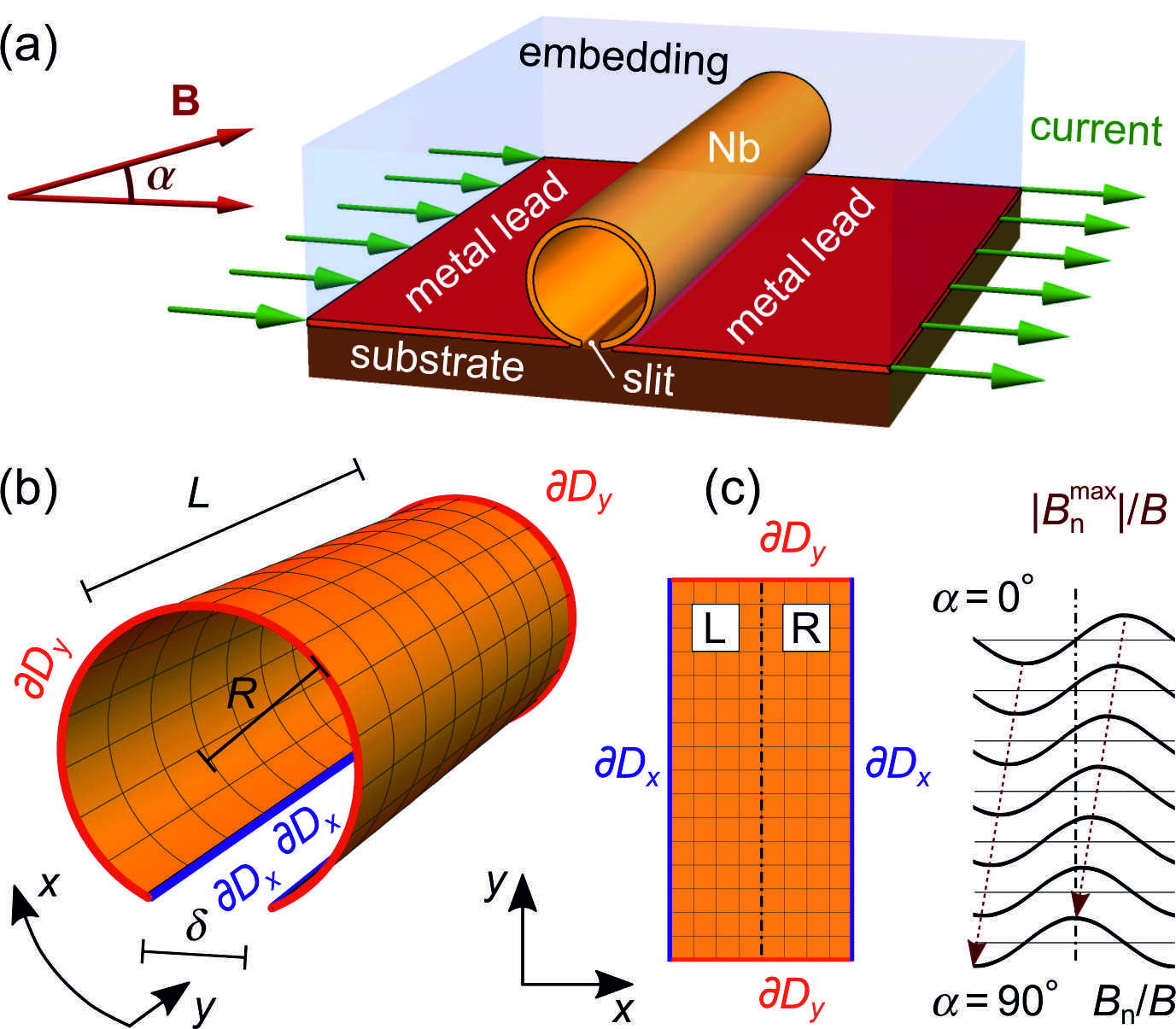}
    \caption{Geometry (a) and the mathematical model (b) of an open nanotube. (c) Unwrapped view of the nanotube surface. Normally conducting current leads are attached to the slit banks and correspond to the $\partial D_x$ boundaries. Indicated is also the evolution of the location of the $|B_\mathrm{n}|$ maxima with increase of the magnetic field tilt angle $\alpha$.}
    \label{f1}
\end{figure}

An extension of a planar film into 3D brings about inhomogeneity\,\cite{Rom07prb,Tem09prb,Gla12prb,Fom22apl} of the normal-to-the-surface component $|B_\mathrm{n}|$ of the applied magnetic field of induction $\mathbf{B}$, such that vortices nucleate and move in the regions where $|B_\mathrm{n}|$ is maximal\,\cite{Fom12nal}. In this regard, of particular interest are open superconductor nanotubes (with a slit), see Fig.\,\ref{f1}, in which, under an azimuthal transport current, vortices are constrained to move within the half-tubes and vortex jets are \emph{non-expanding}\,\cite{Bog23arx}. Previously, correlated dynamics of vortices in open nanotubes were investigated numerically with foci on pinning\,\cite{Bra23arx}, ac drives\,\cite{Fom22nsr}, topological transitions\,\cite{rezaev2020topological}, and trancient regimes\,\cite{bogush2022topological}. These predictions can be examined for, e.g., open Nb nanotubes fabricated by the self-rolling technology\,\cite{Thu08jpd,Thu10nal,Loe19acs}. However, while the average induced voltage $U$ and its frequency spectrum $f_\mathrm{U}$ contain information on the vortex arrangements\,\cite{Bog23arx}, the effects of magnetic-field tilting on their stability and transitions between these arrangements have not been investigated so far.

\begin{figure*}
    \centering
    \includegraphics[width=0.95\linewidth]{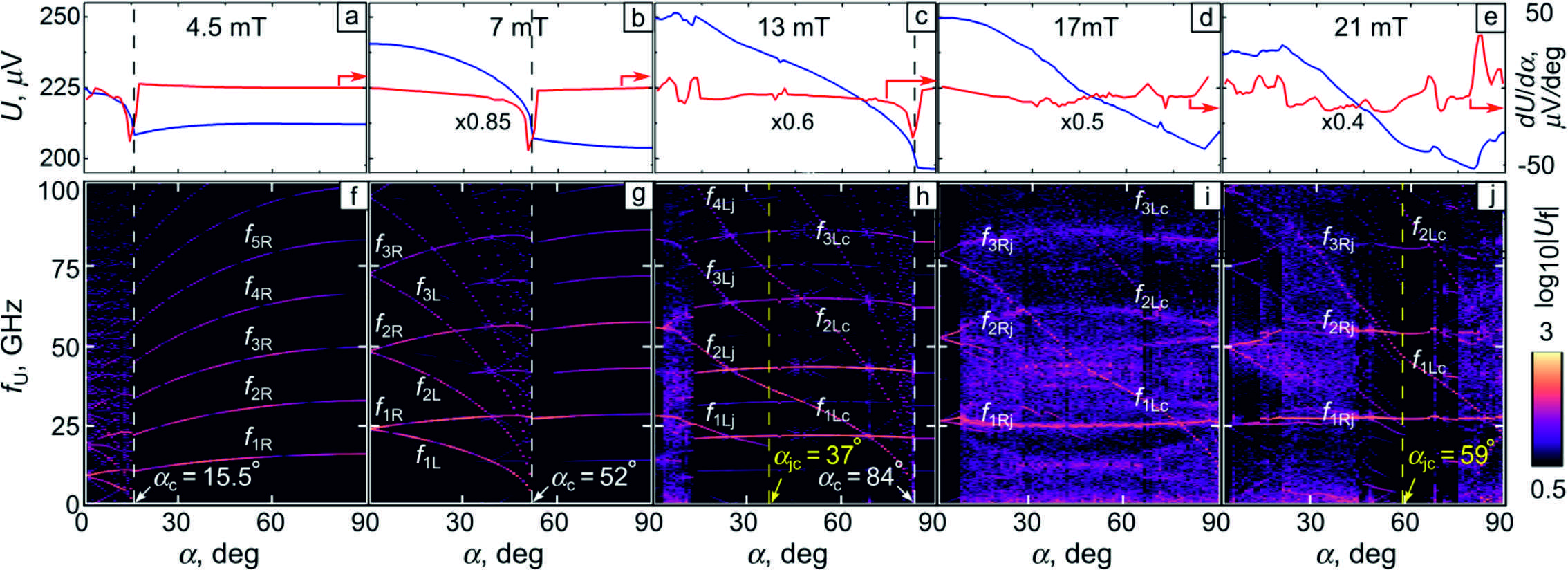}
    \caption{(a-e) Average induced voltage $U$ and (f-j) its frequency spectrum $f_\mathrm{U}$ as functions of the magnetic-field-tilt angle $\alpha$ for the nanotube with $R=390$\,nm at the transport current density $16$\,GA/m$^2$.}
    \label{f2}
\end{figure*}

Here, we present an approach for steering of vortices in open superconductor nanotubes by tilting the vector $\mathbf{B}$ at an angle $\alpha$ in the plane perpendicular to the nanotube axis. The numerical modeling is based on the time-dependent Ginzburg-Landau (TDGL) equation. Distinct from $\alpha=0^\circ$, when vortices move paraxially in opposite directions within each half-tube, an increase of $\alpha$ displaces the areas with close-to-maximum $|B_\mathrm{n}|$ to the close(opposite)-to-slit regions, giving rise to descending (ascending) branches in $f_\mathrm{U}(\alpha)$. At lower $B$, a critical angle $\alpha_\mathrm{c}$ is revealed, at which the close-to-slit vortex chains disappear and $f_\mathrm{U}$ evolves to the $nf_1$-type [$n\geq1$: an integer, $f_1$: vortex nucleation frequency]. At higher $B$, $f_\mathrm{U}$ is largely blurry, due to multifurcations of the vortex trajectories in the opposite-to-slit vortex jet moving in the reverse direction with respect to the close-to-slit vortex chains. In all, our findings have implications for tuning of GHz-frequency spectra in microwave applications and on-demand steering of vortices as information carriers.

The studied geometry is shown in Fig.\,\ref{f1}(a). An open superconductor nanotube of length $L=5\,\mu$m, radius $R=390$\,nm, with a slit of width $\delta=60$\,nm, is exposed to an azimuthal transport current of density $\mathbf{j}_\mathrm{tr}$. A magnetic field of induction $\mathbf{B}$ is applied perpendicular to the tube axis at an angle $\alpha$ relative to the substrate plane, varying between $0^\circ$ and $90^\circ$. Under the action of the transport current, vortices nucleate at the free edges [boundaries $\partial D_y$ in Fig.\,\ref{f1}(b)] of the tube, move along the tube axis, and denucleate at the opposite free edges. At $\alpha=0^\circ$, the vortices in the opposite half-tubes move in reverse directions due to the sign reversal of $B_\mathrm{n}$, see Fig.\,\ref{f1}(c). In what follows, we refer to the vortex arrangements and the half-tubes as ``R'' (right) and ``L'' (left), see Fig.\,\ref{f1}(c). With increase of $\alpha$, the R maximum of $B_\mathrm{n}$ shifts towards the opposite-to-slit region, while the L maximum of $|B_\mathrm{n}|$ shifts towards the L slit bank. This continues until two $|B_\mathrm{n}|$ maxima occur just at the slit banks outside of the nanotube and the $B_\mathrm{n}$ maximum coincides with the middle of the nanotube surface at $\alpha=90^\circ$.

\begin{SCfigure*}
    \centering
    \includegraphics[width=1.68\linewidth]{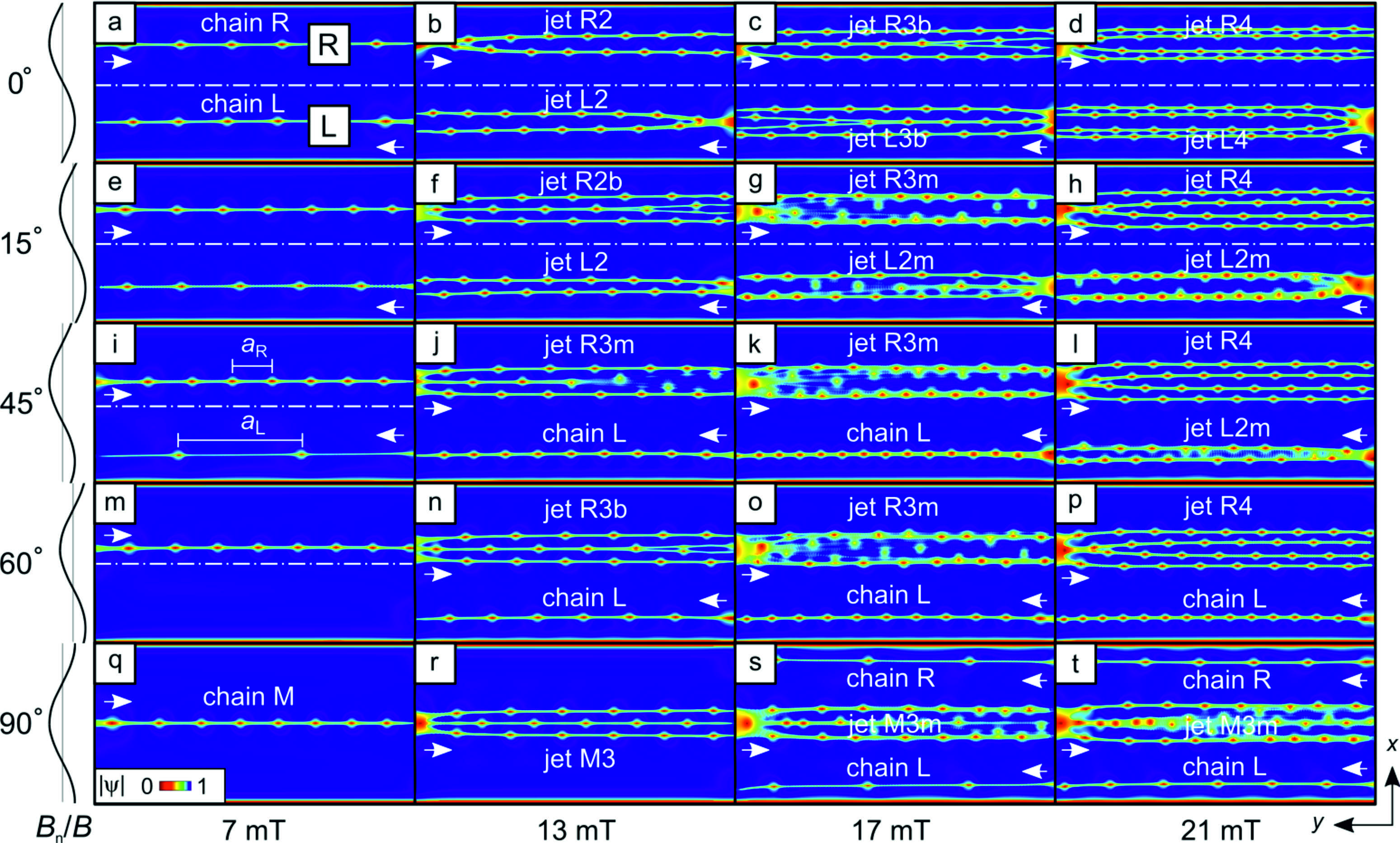}
    \caption{Snapshots of the absolute value of the superconducting order parameter $|\psi|$ overlaid with the accumulated vortex paths for the nanotube with $R=390$\,nm at the transport current density $j_\mathrm{tr} = 16$\,GA/m$^2$. L: left; R: right (half-tube); M: middle; b: bifurcations; m: multifurcations. The number in the jet name corresponds to the number of vortex chains in the jet. Direction of the vortex motion is indicated by the arrows. The dash-dotted lines are the midlines.}
    \label{f3}
\end{SCfigure*}

The TDGL equation is solved for parameters typical for Nb films\,\cite{Dob12tsf} and the film thickness $d = 50$\,nm resulting in the correspondence of a current density of $1$\,GA/m$^2$ to a transport current of $0.25$\,mA. Details on the equations and boundary conditions are provided elsewhere\,\cite{Bog23arx}. The modeling is performed for $j_\mathrm{tr} = 16$\,GA/m$^2$ at temperature $T/T_\mathrm{c} = 0.952$, where $T_\mathrm{c}$ is the superconducting transition temperature.

Figure\,\ref{f2} presents the average induced voltage $U$, its derivative with respect to the magnetic-field-tilt angle $\alpha$, and the induced-voltage frequency spectrum $f_\mathrm{U}$ as functions of $\alpha$ for a series of magnetic field values. At lower fields ($4.5$-$13$\,mT), as $\alpha$ increases, $f_\mathrm{U}(\alpha)$ manifests an abrupt transition from a regime with crossing ascending and descending branches to a regime of $nf_1$-harmonics of the vortex nucleation frequency $f_1$. The transitions occur at some critical angle $\alpha_\mathrm{c}$, which increases with increase of $B$. At the same time, $U(\alpha)$ manifests a sharp bend at $\alpha\lesssim\alpha_\mathrm{c}$, while $dU/d\alpha$ exhibits a minimum at $\alpha_\mathrm{c}$. Whereas $f_\mathrm{U}(\alpha)$ decreases for  $\alpha<\alpha_\mathrm{c}$, it slowly increases for $\alpha>\alpha_\mathrm{c}$ at $\lesssim7$\,mT and is almost constant for $\gtrsim7$\,mT. At higher fields ($17$-$21$\,mT), $f_\mathrm{U}$ is blurry, though one can recognize constant-frequency and descending branches. Overall, $U(\alpha)$ decreases by about $20\%$ as $\alpha$ increases from $0^\circ$ to $90^\circ$.

Further insights into the features of $f_\mathrm{U}(\alpha)$ can be gained from the analysis of the spatial evolution of the absolute value of the superconducting order parameter $|\psi|$ as functions of $\alpha$ and $B$. Indeed, the evolution of $U$ and $f_\mathrm{U}$ at $4.5$\,mT and $7$\,mT can be understood with the aid of the snapshots of $|\psi|(x,y)$ overlaid with the accumulated vortex trajectories in Fig.\,\ref{f3}. Thus, at $7$\,mT and $\alpha=0^\circ$, the vortices are arranged in two chains located symmetrically relative to the (dash-dotted) midline. At $\alpha=0^\circ$, the intervortex distances in the L and R chains are equal, $a_\mathrm{L} = a_\mathrm{R}$. A tilt of $\mathbf{B}$ by $15^\circ$ leads to the inequality $a_\mathrm{R} < a_\mathrm{L}$, which becomes more pronounced with $a_\mathrm{L} \simeq 3a_\mathrm{R}$ at $\alpha=45^\circ$. While the vortex velocities in the R and L half-tubes remain almost equal to each other, the vortex nucleation frequencies in the R and L half-tubes differ by a factor of $f_\mathrm{R}/f_\mathrm{L}\simeq 3$. This ratio agrees well with the frequency ratio $f_\mathrm{1R}/f_\mathrm{1L}$ at $\alpha=45^\circ$ in Fig.\,\ref{f2}(g).

At $7$\,mT, upon reaching the critical angle $\alpha_\mathrm{c} = 52^\circ$, the L vortex chain disappears, decisively affecting the spectrum $f_\mathrm{U}$. Namely, for angles $(\alpha_\mathrm{c} -\alpha)/\alpha_\mathrm{c} \ll 1$, the branches $nf_\mathrm{1L}$ descend very rapidly and $U(\alpha)$ manifests a fast decrease, see Fig.\,\ref{f2}(b). For $\alpha>\alpha_\mathrm{c}$, the spectrum is of the $nf_\mathrm{1R}$-type.  Herein, $f_\mathrm{1R}$ are associated with the harmonics of the vortex nucleation frequency in the single L chain which is displaced to the opposite-to-slit area (midline), see Fig.\,\ref{f3}(q). The vortex arrangements at $4.5$\,mT are qualitatively similar, but differ by a smaller number of vortices and a smaller $\alpha_\mathrm{c}$.

At $13$\,mT and for $\alpha=0^\circ$, there are two vortex jets each consisting of two vortex chains and moving within each half-tube. At $3^\circ\lesssim \alpha \lesssim 13^\circ$, the R chain exhibits multifurcations of vortex trajectories and turns into a jet of three vortex chains, Fig.\,\ref{f3}(f). The multifurcations result in a blurry $f_\mathrm{U}$, Fig.\,\ref{f2}(h). Upon reaching the angle $\alpha_\mathrm{jc} = 37^\circ$, the L jet turns into a vortex chain, Fig.\,\ref{f3}(j). Accordingly, $f_\mathrm{U}$ evolves from the $nf_\mathrm{1Lj}$-type to the $nf_\mathrm{1Lc}$-type, Fig.\,\ref{f2}(h), with $f_\mathrm{1Lc}/f_\mathrm{1Lj}\approx2$ at $\alpha_\mathrm{jc} = 37^\circ$. The descending branches $nf_\mathrm{1Lc}$ reach zero at $\alpha_\mathrm{c} = 84^\circ$, pointing to the absence of an L chain at larger values of $\alpha$, Fig.\,\ref{f3}(r).

At $17$\,mT ($21$\,mT) and for $\alpha=0^\circ$, the jets consist of three (four) chains of vortices, respectively. Their evolution with increase of $\alpha$ can be outlined as follows. The R jet is continuously shifted to the opposite-to-slit region. At $17$\,mT, multifurcations of the vortex trajectories are more pronounced, what makes $f_\mathrm{U}$ more blurred. One can recognize a few descending branches in $f_\mathrm{U}$ in\,Fig.\,\ref{f2}(i), which are due to the reduction of the number of vortices in the L chain as it approaches the L slit bank. Note, for both fields at $\alpha=90^\circ$, a vortex chain appears in the R half-tube, making the vortex pattern symmetric relative to the midline. At $21$\,mT, a transition from the L jet to an L chain occurs at $\alpha_\mathrm{jc} = 59^\circ$, Fig.\,\ref{f3}(p). In this way, with increase of $\alpha$ from $0^\circ$ to $90^\circ$, vortices can be steered to any point of the nanotube, while features in $f_\mathrm{U}$ allow one to attribute them to particular vortex configurations and transitions between them. It is interesting to note, that with increase of $\alpha$, one can realize a transition from a vortex jet to a vortex chain, which does not occur in planar thin films at moderately strong currents (i.e. when non-equilibrium effects can be ignored).

Thus, the major findings can be summarized as follows. First, the symmetry breaking associated with increase of $\alpha$ leads to an increasingly stronger constraint of vortices in the L half-tube, while the primary effect of the magnetic-field tilting on the vortices in the R half-tube consists in their displacement towards the opposite-to-slit area. Accordingly, the frequency of vortex nucleation in the R half-tube remains almost constant, and this is reflected in very slowly increasing or almost constant-frequency branches in $f_\mathrm{U}$. Second, the vortex nucleation frequency in the L half-tube decreases with increase of $\alpha$. At lower magnetic fields ($4.5$--$13$\,mT) this leads to a vanish of the L vortex chain at the critical angle $\alpha_\mathrm{c}(B)$. At moderate and higher fields ($4.5$--$13$\,mT), when the initial configuration of vortices at $\alpha=0^\circ$ evolves into vortex jets in both half-tubes, there is a transition from the L vortex jet to an L vortex chain at the transition angle  $\alpha_\mathrm{jc}$. Both angles $\alpha_\mathrm{c}$ and $\alpha_\mathrm{jc}$ increase with rising $B$. Third, the transition from a vortex jet to a vortex chain or change in the number of vortex chains forming a vortex jet may be not abrupt, but occurring as a consequence of bi- or multifurcations of the vortex trajectories. In this case, the frequency spectrum $f_\mathrm{U}$ is blurry, which complicates the identification of frequencies. Fourth, while \emph{unidirectional} vortex motion via a single vortex chain or a vortex jet in the opposite-to-slit region is revealed for lower fields, a \emph{bidirectional} vortex motion mediated by a centrally located vortex jet and two vortex chains at the slit banks is predicted for higher magnetic fields.

\begin{figure}
    \centering
    \includegraphics[width=0.82\linewidth]{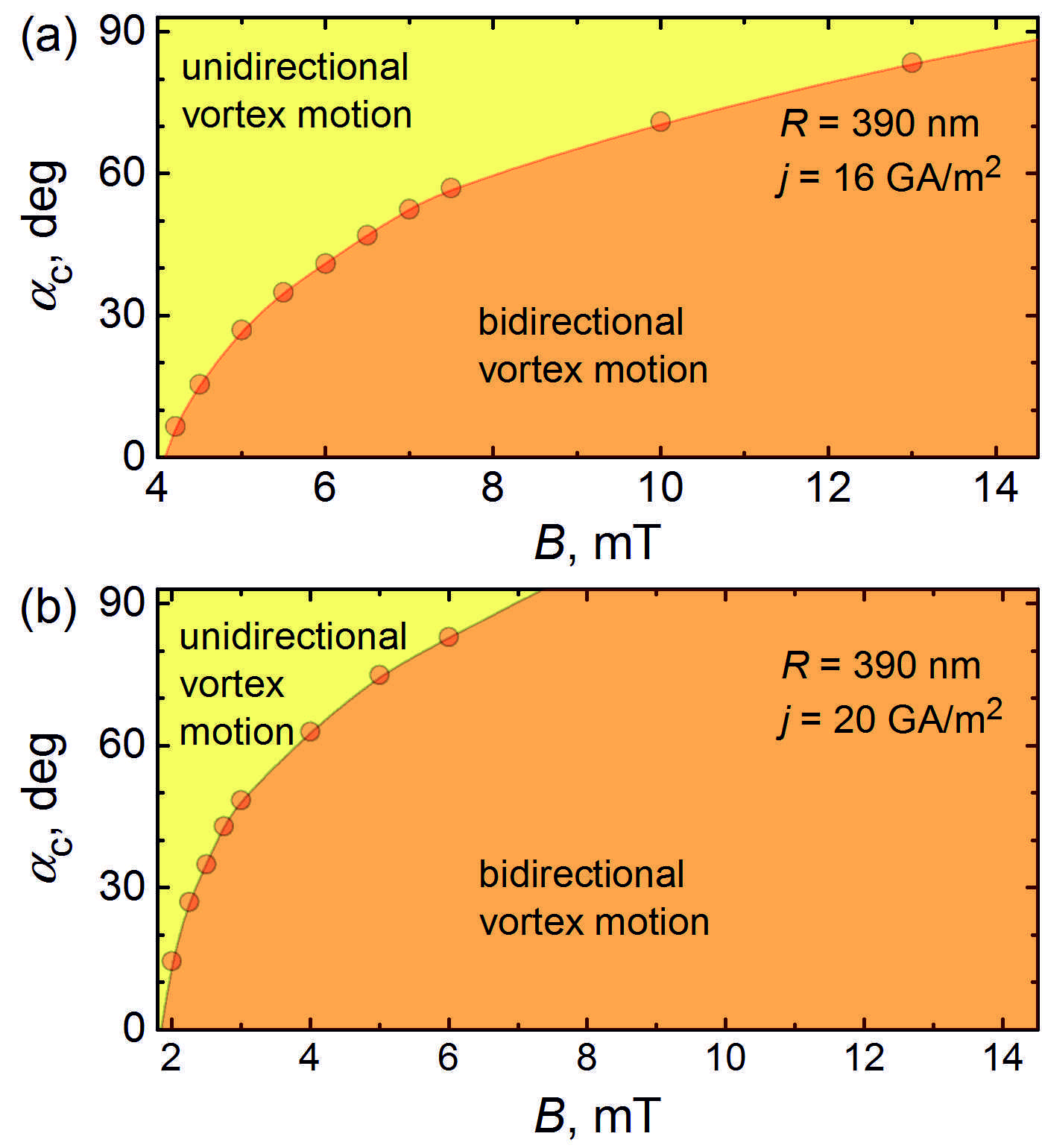}
    \caption{The critical angle $\alpha_\mathrm{c}$ as a function of the magnetic field at $j_\mathrm{tr} = 16$\,GA/m$^2$ (a) and $j_\mathrm{tr} = 20$\,GA/m$^2$(b), demarcating the regimes of unidirectional and bidirectional vortex motion.}
    \label{f4}
\end{figure}

The regimes of unidirectional and bidirectional vortex motion are shown in Fig.\,\ref{f4}, demarcated by the dependences $\alpha_\mathrm{c}(B)$ for two current densities. The shape of the $\alpha_\mathrm{c}(B)$ at $j_\mathrm{tr}=16$\,GA/m$^2$ [see Fig.\,\ref{f4}(a)] implies that the highest sensitivity of the nanotube to the magnetic-field direction is achieved at about $4.25$\,mT. This magnetic field corresponds to the first occurrence of vortices in the nanotube, what makes it very sensitive to the magnetic-field direction. With increase of $j_\mathrm{tr}$ to $20$\,GA/m$^2$ [see Fig.\,\ref{f4}(b)], the boundary between the unidirectional and bidirectional vortex motion regimes shifts towards lower magnetic fields. In addition, from a different study\,\cite{Bog23arx}, it follows that all features in the induced-voltage frequency spectra shift towards higher magnetic fields with decrease of the tube radius.

Finally, we would like to comment on a possible experimental examination and the general relevance of the obtained results. We note that deviations of $\mathbf{B}$ by an angle $\alpha$ of few degrees from the perpendicular-to-tube-axis and parallel-to-substrate direction may easily occur in experiments\,\cite{Loe19acs}, so transitions in $f_\mathrm{U}$ could be useful for checking the sample/field alignment. In addition, the evolution of vortex arrangements with increase of $\alpha$ is interesting from the viewpoint of both, basic research\,\cite{Ols97prb,Gla16prb} and emerging functionalities. For the tubes studied here, it is the steering of vortex arrays to desired parts of the nanotube via magnetic-field tilting and vortex-jet-to-vortex-chain transitions, which do not occur in planar thin films at moderately strong transport currents\,\cite{Bez22prb,Bev23pra}. Conceptually, steering of vortex chains, vortex jets and more complex arrangements is similar to using a magnetic field for controlling the vortex dynamics\,\cite{Vel08mmm}, but surpasses vortex guiding in nanoengineered pinning landscapes in terms of reconfigurability\,\cite{Sta64phl,Sil10inb,Dob20pra}. While spatio- and time-resolved studies of vortex arrangements in 3D nanoarchitectures are challenging, deduction of their properties from features in the global observables $U$ and $f_\mathrm{U}$ represents a viable approach to that end.\\[3mm]

\small{
IB is grateful to Victor Ciobu for technical support, including the generous provision of several servers that were instrumental for the calculations. OVD acknowledges the Austrian Science Fund (FWF) for support through Grant No. I 6079 (FluMag). VMF expresses his thanks to the European Cooperation in Science and Technology for support via Grant E-COST-GRANT-CA21144-d8436ac6-b039a83c-fa29-11ed-9946-0a58a9feac02 and to the ZIH TU Dresden for providing its facilities for high throughput calculations. This article is based upon work supported by the E-COST via Action CA21144 (SuperQuMap).
}


%

\end{document}